# A Dimensional Analysis of the Galactic Baryonic Tully-Fisher Relation with Comparison to MOND and Dark Matter Halo Dynamics

Jeffrey M. La Fortune
1081 N. Lake St. Neenah, WI 54956 forch2@gmail.com
9 April 2020

*Abstract*
Most rotationally-supported galaxies strictly follow the Baryonic Tully-Fisher Relation (BFTR) linking circular velocity with baryon content. This firmly established empirical relationship is currently thought to have origins in either modified gravity or dark matter halo effects. In this work, we construct a physically-based version of the BFTR founded on known scaling relations, disk dynamics (acceleration, jerk and snap) that also reveals the foundational elements responsible for this phenomenology. We employ the Milky Way galaxy as an exemplar to quantitatively compare the two leading theories against this improved version. Additionally, a dimensionless variant of the BFTR is also provided which may permit its use as an analytic tool to aid in the understanding of galactic dynamics.

*Introduction*
In this paper we examine the BFTR through the lens of dimensional analysis and application of physical properties attributed to galactic disks; baryons, energy and angular momentum (La Fortune 2019). We quantify the physical nature of galactic disks via acceleration, jerk and snap parameters attributed to rotationally supported, distributed mass systems. Employing the Milky Way as an exemplar, we provide a detailed and quantified component break-down not possible with MOND (Milgrom 1983a) (Milgrom 1983b) and modeled dark matter halo (McGaugh 2018). We demonstrate the improved utility and precision of this reformulated BTFR compared to current interpretations.

*General MOND Theory*
The simplified MOND BFTR equation includes a characteristic or universal acceleration ($a_0$) in addition to G, the gravitational constant:

$$V_{flat}^4 / GM_{Bar} \cong a_0$$

In the above equation, the right-hand term is Milgrom's Constant, an empirically established fundamental acceleration value that provides "best fit" rotation profiles for an entire class of rotationally supported galaxies. This constant is entirely contingent on galactic phenomena; the observed velocity – baryon ratios. A dimensional form of the MOND BTFR is shown below:

$$V_{flat}^4 = \left(\frac{1.2 x 10^{-13} km}{s^2}\right)\left(\frac{1.33 x 10^{11} km^3}{M_{Bar} s^2}\right) M_B = \left(\frac{0.01594 km^4}{M_{Bar} s^4}\right) M_B$$

$$V_{flat}^4 = a_0 G M_{Bar} = 0.01594 M_{Bar}$$

For the balance of this analysis, the MOND flat velocity is also taken to be the scaling circular velocity ($V_{flat} = V_{Circ} = V_\infty$). Milgrom's constant is given as $a_0 = 1.2 x 10^{-13}$ kms$^{-2}$ and the External Field Effect (EFE) is not included. The regions of the galaxies we are interested are in the Deep MOND Limit of exceedingly low acceleration.

*MOND and Dark Matter Halo Models of the Milky Way*

We employ McGaugh's updated Q4MB model of the Milky Way galaxy with a baryon content $M_{Bar}$ =7.38x10$^{10}$M$_\odot$ (McGaugh 2018). We selected this model as it employs the empirical RAR and the observed rotation curve to map inner disk baryonic surface density to the solar position with good accuracy. McGaugh also fits the data to a pseudo-isothermal dark matter halo with highly constrained core radius $R_C$ =3.1 kpc and an 'asymptotic velocity' $V_\infty$=185.8 kms$^{-1}$. The total mass of McGaugh's 'implied' halo is <10$^{12}$M$_\odot$. This mass value agrees with latest estimates obtained from Gaia satellite galaxy kinematics (Fritz 2020). In this study, Fritz estimated dynamic mass <64 kpc to be 0.58x10$^{12}$M$_\odot$ [span - 0.44 to 0.63] with a virial mass 1.51 x10$^{12}$M$_\odot$ [span – 1.11 to 1.96]. This is consistent with the scaling model with a dynamic mass $M_{Dyn}$ =0.5x10$^{12}$M$_\odot$ and a virial mass twice that value. In practical terms, McGaugh indirectly links MOND with dark matter halo effects representing the global dynamical properties of the Milky Way galaxy.

Recently, it has been demonstrated that fuzzy dark matter bridges the divide between the two motivations with all models reproducing two phenomenologically significant galactic scaling relations, the Radial Acceleration Relation (RAR) and the BTFR (Lee 2019). This proposal demonstrates that the action of dark matter halos can be made dynamically indistinguishable to that of modified gravity. From many different avenues, an established motivation for the origin of these scaling relations is still being sought.

We begin derivation of the scaling BTFR by calculating the circular velocity of the Milky Way via MOND using McGaugh's baryon mass. We find this velocity agrees with the simple MOND BTFR expectation:

$$V_{Circ} = 185.5 \, kms^{-1} = \frac{\sqrt[4]{M_{Bar}}}{2.81} = \frac{\sqrt[4]{0.0738 x 10^{12}}}{2.81} \; ; \quad \frac{1}{\sqrt[4]{0.01594}} = 2.81$$

MOND also establishes a dimensionless mass ratio termed the mass discrepancy factor, D linking galactic dynamic mass to baryonic mass:

$$D = \frac{M_{Dyn}}{M_{Bar}} = \frac{V_{obs}^2}{V_{Bar}^2}$$

$M_{Bar}$ is the gravitating baryonic "rest mass." When *D=1*, the observed velocity is equal to the baryon contribution and for a point mass equivalent, $V_{Bar} = \sqrt{GM_{Bar}/R}$. We show that mass discrepancy is an important parameter in defining the dynamic nature of the galactic disks.

Based on the MOND definition of the BTFR and a baryonic point mass, we make a substitution in observed velocity term to obtain the effective D for the McGaugh model:

$$D = \frac{V_{obs}^2}{V_{Bar}^2} = \frac{(\sqrt[4]{M_{Bar}}/2.81)^2}{(\sqrt{GM_{Bar}/R})^2} = \frac{R_M}{9.22} = \frac{62.4}{9.22} = 6.77$$

With flat rotation velocity $V_{Circ}$ =185.5kms$^{-1}$ and mass discrepancy D=6.77, the corresponding disk radius is $R_M = 62.4$ kpc (for $M_{Dyn} = 0.5$x10$^{12}$M$_\odot$). These parameters describe the scaling equivalent (ansatz) of McGaugh's Milky Way galaxy model.

In Figure 1 the McGaugh model is illustrated with a log-log 'Radius-Velocity' plot for r >10 kpc where MOND dominates. The circular velocity is constant (horizontal blue dash) within and beyond the disk's edge ($R_M$) to indeterminate radii. Central point mass baryon velocity support is shown for $M_{Bar}$ (black solid). The mass discrepancy function (red solid) is shown on right hand axis. It continues to steeply rise beyond 1000 kpc.

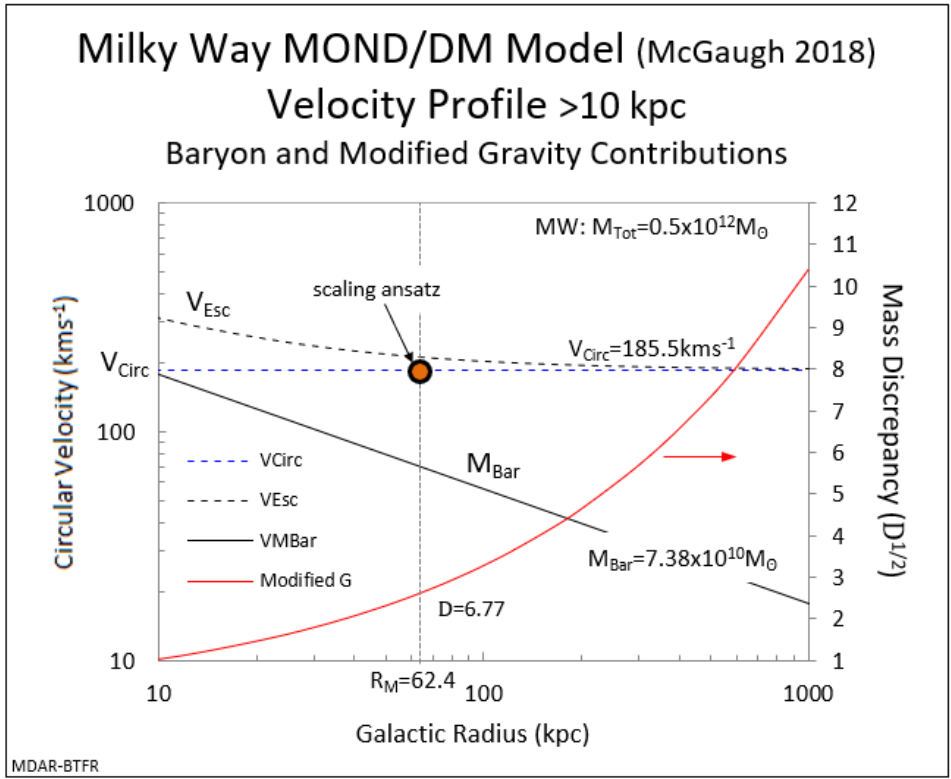

*Figure 1: The McGaugh model of the Milky Way based on MOND precepts and target dynamic mass $0.5 \times 10^{12} M_\odot$ (orange filled black circle). This model illustrates velocity support for point mass baryons (black solid), modified gravity effect (red solid), and baryon sourced velocity dispersion (black dashed) as a function of Galactic radius.*

We make a particular point to discuss the MOND escape velocity ($V_{Esc}$) profile (black dash). This profile is exclusively associated with baryonic mass only as modified gravity has exhibited no capacity to "thermalize" matter in the conventional sense. In the scaling model, thermodynamic processes and effects are included in the total dynamic.

From the perspective illustrated in Figure 1, escape velocity decreases with distance while circular velocity remains constant with the differential between $V_{Esc}$ and $V_{Circ}$ narrowing with radius. This sets the stage for cold, circular orbits around the Milky Way that have not been observed. At large radii, MOND supported orbits should become progressively susceptible to destabilization. Also, there is another concern as these dynamics prohibit formation of a homogeneous and isotropic universe although this is still a topic of debate (Felton 1984).

In order to provide a more reasonable physical interpretation of MONDian escape velocities in low acceleration regime, Milgrom considered the energy associated with the logarithmic potential required to support highly extended circular velocities (Milgrom 2014).

This treatment results in an infinite potential – in other words, a truly 'Machian' universe. Presented in either form, the MOND $V_{Esc}$ profile has not been confirmed via observation and remains unresolved. The scaling proposal eliminates this uncertainty by defining the physical edge of the baryonic disk and respecting Keplerian dynamics beyond the edge.

*General Scaling Model*
Much work has been conducted examining the phenomenological scaling relations of rotationally-supported galaxies. Many studies have revealed simple power law relationships between three fundamental galactic properties; $M_{Dyn}$, $V_{Circ}$ and $R_{Disk}$. Rather than invoking modified gravity or dark matter halos, we consider the gravitating potential of highly ordered disk energy as the source of galactic 'missing mass.'

Much emphasis is placed on empirical constraints as guides to obtain the fundamental scaling parameters and relations. These scaling relations are found to be highly dependent on dynamic mass surface density as measured from the edge of the disk ($\mu_{RD}=M_{Dyn}/\pi R_{Disk}^2$). We find for a particular galactic mass surface density:

$$\mu_{MRV} = \frac{M_{Dyn}}{\pi R_D^2} = 67 M_\odot pc^{-2},$$

the scaling 'TFR' relation simplifies to:

$$V_{Circ} = \frac{\sqrt[4]{M_{Dyn}}}{4}$$

We term any galaxy with $\mu_{RD} = \mu_{MRV}$ as "being on" the [$M_{Dyn}$-R-V] relation per the above equation, the dynamic 'TFR' applies explicitly. The relation imposes strict 'm=4' power law functionality, similar to MOND. This scaling equation is in agreement with an earlier study of 118 spiral and irregular galaxies sampled from the Spitzer Photometry & Accurate Rotation Curve (SPARC) survey (Lelli 2016). His observations establish the power law relationship with low intrinsic scatter. As demonstrated above, the residuals showed no correlation to disk morphology, surface density, radius, or gas fraction.

The [$M_{Dyn}$-R-V] relation is only applicable for galaxies having a dynamic surface density $\mu_{MRV}$ representing a small subset of the galactic population. Since all galaxies with similar dynamic mass do not have identical circular velocities, a second disk parameter must be introduced – observed dynamic mass surface density $\mu_{RD}=M_{Dyn}/\pi R_D^2$. By including with parameter as the ratio between $\mu_{RD}$ and $\mu_{MRV}$, the scaling 'TFR' accommodates any physical disk density providing the adjusted velocity. With this parameter, the generalized scaling BTFR becomes:

$$V_{Circ} = \frac{\sqrt[4]{D(\mu_{RD}/\mu_{MRV})M_{Bar}}}{4}$$

For a given $M_{Dyn}$, a highly diverse population of rotation velocities are possible (as both $\mu_{RD}$ can D can vary). This increased flexibility and precision not possible for MOND utilizing constant pre-factor "$a_0 G$."

*The Scaling Model for the Milky Way Galaxy*

We employ the [$M_D$-R-V] relation to determine the Milky Way's rotation velocity with $M_{Dyn}$ = 0.5x10$^{12}$M$_\odot$ and D=5.9. Although the Galaxy is not precisely on this relation, we use it for demonstration purposes:

$$V_{Circ} = \frac{\sqrt[4]{DM_{Bar}}}{256} = \frac{\sqrt[4]{M_{Bar}}}{2.57} = \frac{\sqrt[4]{0.085 x 10^{12}}}{2.57} = 210.1 \ kms^{-1}$$

For a fixed surface density, galaxies with higher mass discrepancies will have higher circular velocities. Since D (and $\mu_{RD}$) each only span one decade in range, velocities are physically constrained for a given $M_{Bar}$. We find that the denominator is the only difference between the [$M_D$-R-V] solution and MOND.

Figure 2 represents the $R_D$-$V_C$ scaling model parameters for the Milky Way on the [$M_D$-R-V] relation (purple solid point). The velocity support for various components are given in the key. Scaling parameters for the Milky Way are $\mu_{RD}$=67 M$_\odot$pc$^{-2}$, $V_C$=210.1 kms$^{-1}$, and $R_D$=48.6 kpc for D=5.9.

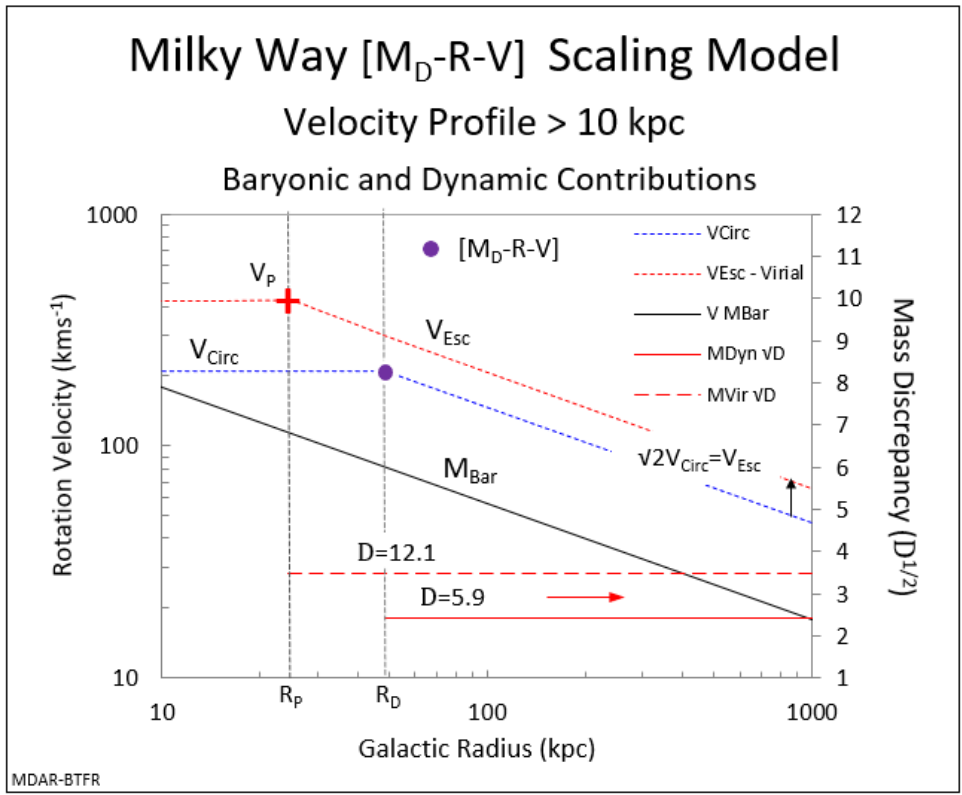

*Figure 2: Milky Way Scaling Model – Rotation velocity versus galactic radius. The [$M_D$-R-V] model is denoted at $V_{Circ}$-$R_D$ (purple solid). The virial $V_P$-$R_P$ (red cross) establishes the global escape velocity profile (red dash) which coincides with the escape velocity profile associated with dynamic disk mass as $V_{Esc}$=√2$V_{Circ}$ >$R_D$ (vertical black arrow). Mass discrepancies are shown on the right-hand vertical axis (horizontal red arrow).*

In Figure 2, we depict radii >10 kpc as the Deep Mond Limit. The viral velocity at $R_P$ is 420.2 kms$^{-1}$, twice $V_C$. The Milky Way's circular velocity (blue dash) is 'flat' to $R_D$ (purple point). The velocity then declines in classical (Keplerian) fashion beyond $R_D$.

Peak virial velocity $V_P$ (red-cross) characterizes the 'global' escape velocity $V_{Esc}$ profile (red dash), equivalent to D=12.1. Unlike MOND mass discrepancy that constantly increased with distance from the Galaxy, scaling has D constant outside $R_D$ inferring a simple disk-sized compact enclosed mass.

Figure 2 also illustrates a unique geometric configuration for galaxies on the [$M_D$-R-V] relation. In this particular configuration, the global escape velocity (red dash beyond $R_P$) coincides with the escape velocity √2 $V_C$, matching the same $V_{Esc}$ profile beyond $R_D$. Although not a unique combination, it is one that may exhibit enhanced thermodynamically stability over other geometric configurations off the [$M_D$-R-V] relation.

*Relational aspects Between MOND $\mathcal{A}_0$=a₀G and Proposed Scaling Model*
In this section we derive an alternate approach to the BTFR based on the dimensional assignment of G and the role acceleration plays with regard to galactic dynamics. A scaling analog to the MOND BTFR is constructed that permits additional parameters and clarifies the meaning of $a_0$ and hence $\mathcal{A}_0$. As depicted in Figure 2, the governing scaling relation is:

$$V_{Circ} = \frac{\sqrt[4]{M_{Dyn}}}{4} \rightarrow V_{Circ}^4 = \frac{M_{Dyn}}{4^4}$$

Before proceeding, we confirm the scaling derived BFTR zero-point is in agreement with the empirical value using above equation:
$$M_{Bar} = \left(\frac{256}{5.9}\right) V_{Circ}^4 = 43 V_{Circ}^4$$

We find this result is physically consistent with $M_{Bar}$ = 47±6$V^4_{flat}$ observed for wide sample of disk galaxies (McGaugh 2012). As with MOND's $\mathcal{A}_0$=a₀G pre-factor, a corresponding scaling acceleration $a_s$ can also be generated that serves a role similar to $a_0$:

$$a_0 G M_{Bar} \approx \frac{D M_{Bar}}{256} \text{ with } a_s = \frac{D}{256G}$$

With this replacement, the characteristic acceleration constant $a_s$ can now expressed as the ratio between D and G. This opens up a physically-based solution space not possible within the MONDian framework. Rearranging, the scaling version of the BTFR incorporating gravitational *constant* G becomes:
$$V_C^4 = a_s G M_{Bar}$$

Solving, the scaling acceleration is $a_s$=1.73x10⁻¹³kms⁻². Unlike the MOND pre-factor (a₀G=0.01594) that suffices for all galaxies, this particular scaling pre-factor (a$_s$G=0.02300) applies to a very thin slice of the galactic population. Figure 3 illustrates the correlation between D and $\mu_{RD}$ for galaxies that fall precisely on the [$M_D$-R-V] relation (diagonal purple dash). Milky Way model parameters are indicated; McGaugh MOND (orange filled black circle) and the scaling model (purple solid).

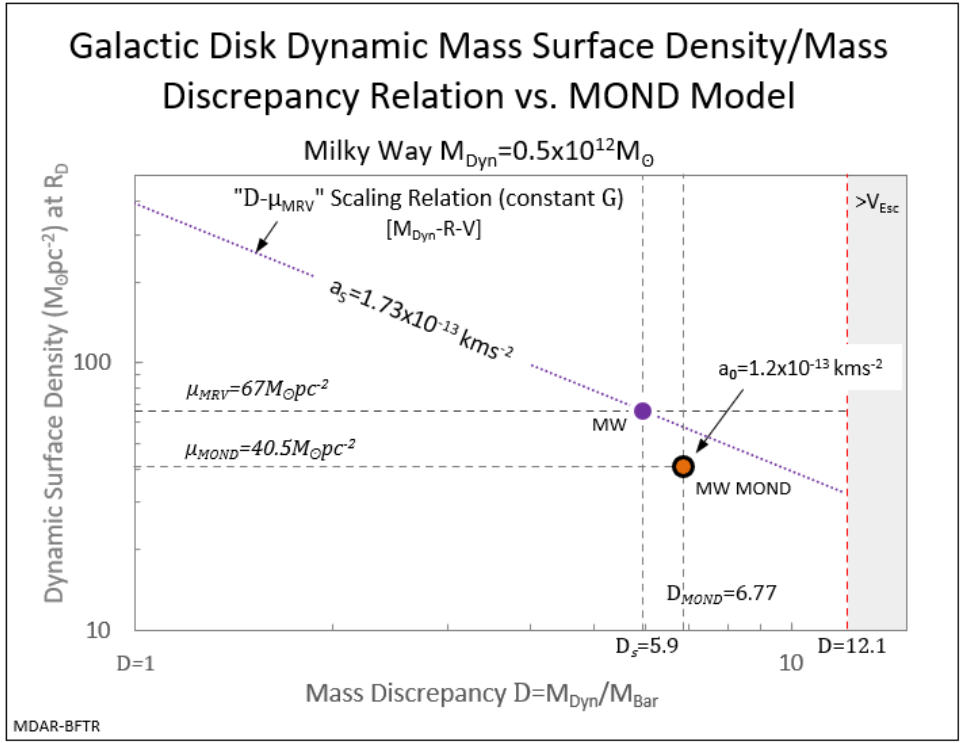

*Figure 3: Galactic dynamic mass surface density versus Mass Discrepancy for the MOND and Scaling Milky Way models. The difference in disk properties between the [$M_D$-R-V] (purple solid) and MOND (orange filled black circle) is gauged by their separation on the plot. Beyond $D=12.1$ (shaded region) global escape velocities are exceeded.*

At first glance Figure 3 appears unassuming but illustrates the universal trend between galactic surface mass density and mass discrepancy. The thermodynamically "preferred D-$\mu_{RD}$ configuration" occurs along the [$M_D$-R-V] relation (purple dash), maximizing thermodynamic stability. We expect massive ellipticals and high brightness galaxies to populate the left side of the plot, classical spirals in the central portion, and low surface brightness spirals, dwarfs, and irregulars towards the right. A significant difference in models is evident with the MOND ansatz having a mass surface density forty percent lower than the [$M_D$-R-V] expectation and sixty-percent lower than the observed Milky Way density per Table 1. The MOND BTFR requires modification of galactic physical parameters to accurately fit the rotation curve. Scaling employs observed galactic physical properties with no need for modification.

*Dimensional Transformation of the MOND BTFR*
In MOND, the BFTR for all rotationally supported galaxies is dimensionally satisfied via $\mathcal{A}_0$ as the product of two constants, $a_0$ and $G$. This simplicity is a positive feature of MOND - one free parameter "$a_0$" and an interpolating function between the low and high acceleration regimes. A dimensional breakdown for the MOND term $\mathcal{A}_0 = a_0 G$ is shown:

$$V^4 = \left(\frac{1.2 \times 10^{-13} km}{s^2}\right)\left(\frac{1.33 \times 10^{11} km^3}{M_{\odot Bar} s^2}\right) M_B = \left(\frac{0.01594 km^4}{M_{\odot Bar} s^4}\right) M_{\odot Bar}$$

Constants →        $a_0$              $G$

This BTFR formulation employs two 'hard' constants $a_0$ and $G$ which we will show is overly constrained.

To establish a more flexible and physical set of parameters, we find the dimensional assignment of G (acceleration divided by a mass surface density) provides a crucial key (Christodoulou 2018) (Christodoulou 2019).

Rather than keeping each constant dimensionally separate, we combine the 'acceleration' from G and the $a_0$ (or $a_s$) term into the simple product ($\alpha^2$) divided by disk dynamic mass surface density $\mu_{RD}=M_{Dyn}/(\pi R_D^2)$ :

$$V^4 = \left[\left(\frac{km^2}{DM_{Bar}}\right)\left(\frac{km}{s^2}\right)\right]\left(\frac{km}{s^2}\right) M_B$$

Variables → $\quad \mu_{RD}^{-1} \quad\quad\quad \alpha^2$

This regrouping permits $\mathcal{A}_0$ (a product of two constants) to become a function of two interdependent variables. For galaxies on the [$M_D$-R-V], the BTFR is recast:

$$V_C^4 = \frac{\alpha^2}{\mu_{MRV}} M_{Bar}$$

As with the previous scaling BTFR, this regrouping accommodates all galactic surface densities and circular rotation curves via a simple ratio with the 'dimensionless' form on the right.:

$$V_C^4 = \frac{\alpha(\mu_{RD})^2}{\mu_{MRV}} M_{Bar} = \frac{D\mu_{RD}}{256\mu_{MRV}} M_{Bar}$$

Although highly simplified, this dimensionless format has practical utility when comparing galactic disk properties. A brief example is shown for the three galaxies listed in the table below:

*Table 1: Example Galaxies: Observed Properties*

| Galaxy | | $M_{Dyn}$ | $R_{Disk}$ | $V_{Circ}$ | $V_{Peak}$ [a] | $\mu_{RD}$ [b] | D [c] |
|---|---|---|---|---|---|---|---|
| Name | Type | x$10^{12}M_\odot$ | kpc | kms$^{-1}$ | kms$^{-1}$ | $M_\odot pc^{-2}$ | -- |
| And IV | VLSB dIrr | 0.0034 | 7.5 | 45 | 45 | 19 | 9.1 |
| Milky Way | HSB | 0.50 | 40 | 230 | 432 | 99 | 5.9 |
| M31 | HSB | 1.47 | 100 | 251 | 467 | 47 | 5.9 |

[a] $V_P$ @ 0.5$R_D$  [b] $M_{Dyn}/\pi(R_D)^2$  [c] $M_{Dyn}/M_{Bar}$

With observed galactic parameters, the dimensionless format accurately calculates circular velocities based on a minimal set of observations or modeling values. For example, the MW:

$$V_C = \sqrt[4]{\frac{D\mu_{RD}}{256\mu_{MRV}} M_{Bar}} = \sqrt[4]{\frac{5.9\,(99)}{256\,(67)}(0.085x10^{12}M_{Bar})} = 232\ kms^{-1}$$

Likewise, circular velocities calculated for And IV and M31 are 44 and 252 kms$^{-1}$, respectively.

As a cross-check, we use McGaugh's parameters (scaling ansatz) and find a near identical result ($V_C$=185.3 vs. 185.5 kms$^{-1}$). More important than the increased accuracy is the elimination of an ad hoc universal acceleration constant and an acceleration squared term permitting higher order time derivatives, jerk and snap to be calculated for the disk.

In Figure 4, the scaling 'space' between dynamic surface density, disk acceleration, and mass discrepancy is constructed. As with the previous figure, the Milky Way model is on the [$M_D$-R-V] relation (purple data point on purple diagonal) and MOND/dark matter model (orange/black circle). Also included for perspective are observed values for And IV (black triangle), MW (open red circle), and M31 (open black circle) per Table 1.

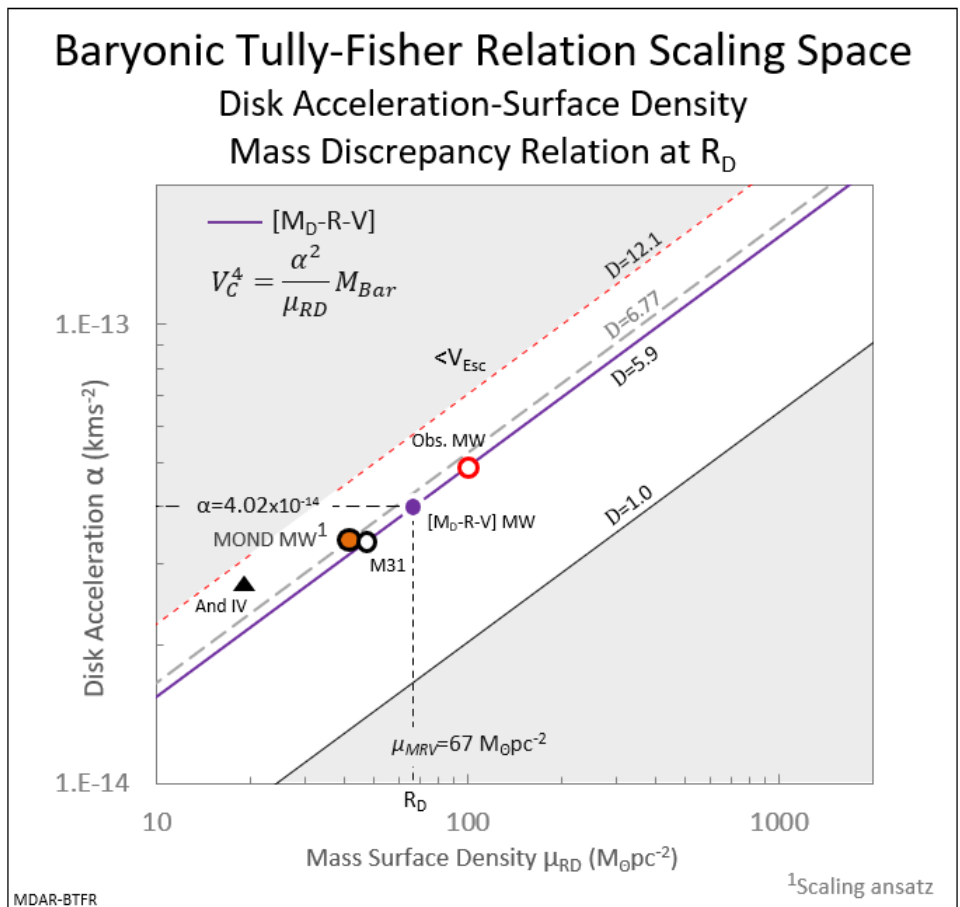

Figure 4:  The complete BTFR scaling model templated as disk edge acceleration versus dynamic mass surface density with accommodation for mass discrepancy. The [$M_D$-R-V] relation (purple solid) runs along D=5.9. The MW scaling model is positioned at $\mu_{MRV}$ (purple solid) with McGaugh's MOND/DM model at D=6.77 (orange filled black circle). Observed galactic parameters for the MW, And IV and M31 are in Table 1. 'Unphysical' galactic properties are outside D=1 and 12.1 (gray shade).

In Figure 4, the scaling BTFR solution space and physical boundaries are depicted. The mass discrepancy axis is bounded above by galactic escape velocity D=12.1 (red dash) and below by D=1 (black solid). As shown in Table 1, the MW and M31 mass discrepancies have been fixed to D=5.9, but can take on any reasonable value (for example, And IV observed D=9.1).

The MOND ansatz for the Milky Way is constrained to a single point in this space, versus the wide array of disk parameters that are physically permissible with the scaling model.

To this point, we have focused on 'global' galactic disk dynamics as represented by the BTFR. In Figure 5, we demonstrate in more detail how this scaling approach is applied to internal disk ($\leq R_D$) properties as well. Below, we present two scaling models, $\alpha^2/\mu(r)$ and $a_S$ (G constant) with a comparison to the MOND ansatz. The virial peak (red cross and diagonal red dash) is at the disk's mid-point ($R_P=0.5 R_D$) and has a dynamic disk mass surface density $\mu_{RP}=134$ M$_\odot$pc$^{-2}$ per the [$M_D$-R-V] relation.

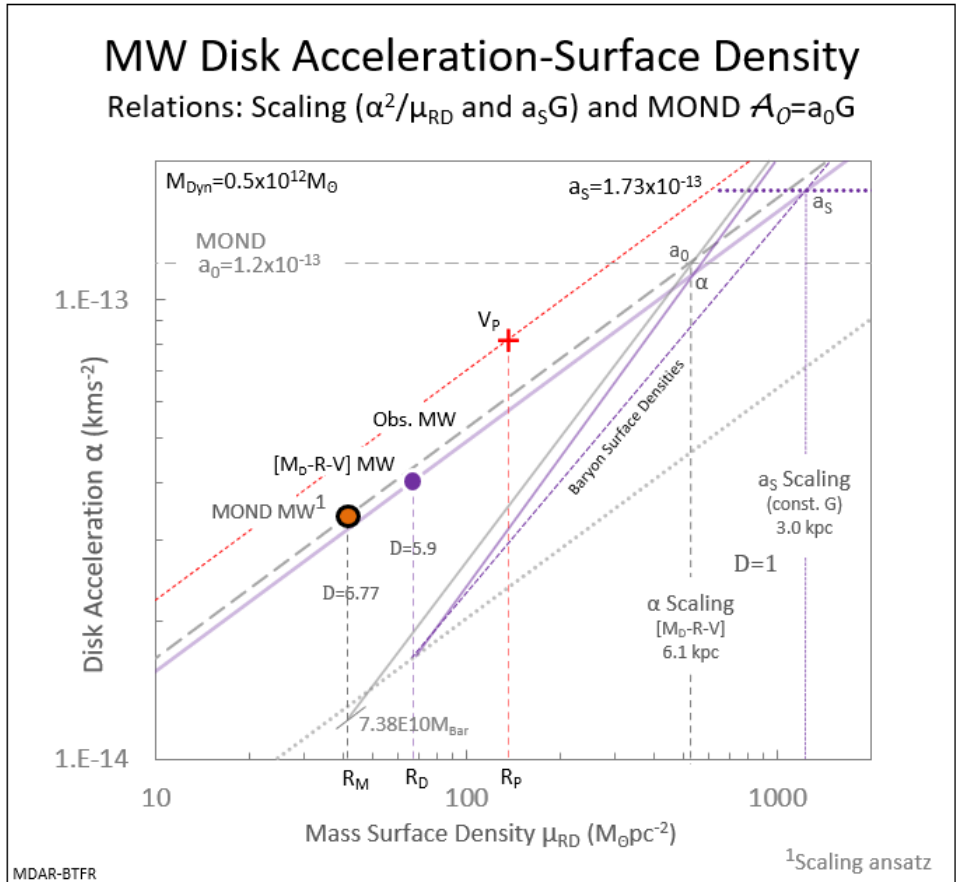

Figure 5: Two Milky Way scaling models (purple) versus McGaugh MOND/DM model (gray). In addition to disk acceleration and dynamic surface density, a third parameter – mass discrepancy is also plotted. There are dynamic similarities between the scaling "α" (solid purple) and MOND "$a_0$" (dashed/solid gray) MOND models. Although MOND's dynamic profile is proven to be accurate, it requires modification of the Milky Way's physical parameters. The discrepancy between the MOND and scaling "α" model properties are due to MOND's relatively low rotation velocity and low dynamic mass surface density compared to observation.

As modeled, the dynamics of the $\alpha^2/\mu$ and $\mathcal{A}_0=a_0G$ relations (solid light purple and gray dash, respectively) are very similar. For the Milky Way, this demonstrates MOND's capability to realistically estimate galactic circular velocities within its capacity. Since scaling dynamics are sourced from baryonic mass, these density profiles are also depicted (steep purple curves labeled $a_0$, $a_S$, and α).

*Higher Order Acceleration Time Derivatives and Interpretation*

The acceleration squared term is very flexible as it can be redimensioned into galactic Jerk and Snap parameters as the first and second time-derivatives of $\alpha$, respectively. In this late-universe era, there is little evolution in these parameters as galaxies are in a state of quiescent quasi-equilibrium. The starting point is the dimensional form of the scaling BTFR:

$$V_C^4 = D \frac{ML^6}{L^2 M^2 T^4}(M_{Bar}) = D\left(\frac{L^2}{M}\right)\left(\frac{L^2}{T^4}\right)(M_{Bar}) = \frac{\alpha^2}{\mu_{RD}}(M_{Bar})$$

The $\alpha^2/\mu_{RD}$ term is flexible and can be rearranged to obtain galactic disk jerk and snap components. The Jerk is regrouped as $j=(L/T^3)$ and a complementary velocity $V=(L/T)$:

$$V_C^4 = D \frac{ML^6}{L^2 M^2 T^4}(M_{Bar}) = D\left(\frac{L^2}{M}\right)\left(\frac{L}{T}\right)\left(\frac{L}{T^3}\right)(M_{Bar}) = \frac{\alpha^2}{\mu_{RD}}(M_{Bar})$$

For $M_{Dyn}=R_D V_C^2/G$, the scaling BTFR is recast into a new dimensional form with a jerk term:

$$V_C^4 = D\underbrace{\left(\frac{km^2}{M_{\odot Bar}}\right)}_{\mu_{MRV}}\underbrace{\left(\frac{km}{s}\right)}_{\text{Velocity}}\underbrace{\left(\frac{km}{s^3}\right)}_{\text{Jerk}}\underbrace{M_{\odot Bar}}_{\text{Mass}} = 0.02300\left(\frac{km^4}{M_{\odot Bar} s^4}\right)M_{\odot Bar}$$

<div style="text-align:center">Mass Disc. Surf. Den.   Velocity   Jerk   Mass</div>

Conservatively taking the velocity term $V_C=V_{Circ}$, jerk is a function of circular velocity cubed. For a galactic disk on the [$M_D$-R-V] relation and dynamic mass $M_{Dyn}=0.5 \times 10^{12} M_\odot$, galactic jerk $j= 3.3 \times 10^{-28}$ kms$^{-3}$. Going back to the BTFR, a second time derivative of disk acceleration is obtained, snap with dimensions $s=(L/T^4)$:

$$V_C^4 = D\left(\frac{L^3}{M}\right)\left(\frac{L}{T^4}\right)(M_{Bar}) = \frac{\alpha^2}{\mu_{RD}}(M_{Bar})$$

We find circular velocity is dependent on galactic volumetric dynamic mass density with snap for a given $M_{Bar}$:

$$V_C^4 = D\underbrace{\left(\frac{km^3}{M_{\odot Bar}}\right)}_{\rho RD}\underbrace{\left(\frac{km}{s^4}\right)}_{\text{Snap}}\underbrace{M_{\odot Bar}}_{\text{Mass}} = 0.02300\left(\frac{km^4}{M_{\odot Bar} s^4}\right)M_{\odot Bar}$$

<div style="text-align:center">Mass Disc. Vol. Den.   Snap   Mass</div>

The snap of the Milky Way's disk is $s = 3.5 \times 10^{-44}$ kms$^{-4}$. Combining, the reduced equation for $V_{Circ}$ becomes a function of disk radius and the ratio between snap and jerk:

$$\text{Jerk: } V_C^3 = \frac{j}{\mu_{RD}}M_{Bar}; \quad \text{Snap: } V_C^4 = \frac{s}{\rho_{RD}}M_{Bar}, \quad \text{with } V_C = \frac{s\,\mu_{RD}}{j\,\rho_{RD}}$$

Plugging in scaling values for the Milky Way, the circular velocity is recovered:

$$V_C = 210.1\, kms^{-1} = \frac{s\mu_{RD}}{j\rho_{RD}} = \frac{(3.5 \times 10^{-44} kms^{-4})(7.01 \times 10^{-26} M_{Bar} km^{-2})}{(3.3 \times 10^{-28} kms^{-3})(3.5 \times 10^{-44} M_{Bar} km^{-3})}$$

Snap and mass volume density numerically cancel and provide correct dimensionality for velocity.

*Milky Way Disk Deceleration Parameter*

As a consistency check for the scaling acceleration, jerk and snap results, the Milky Way's disk deceleration parameter ($q$) at $R_D$ can be calculated via the conventional definition:

$$q_{MW} = -\left(\frac{\ddot{a}a}{\dot{a}^2}\right) = -\left(\frac{s\alpha}{j^2}\right) = -\frac{(3.5 \times 10^{-44} \, kms^{-4})(4 \times 10^{-14} \, kms^{-2})}{(3.3 \times 10^{-28} \, kms^{-3})^2} = -0.01$$

This result in in agreement with $q \equiv 0$ for a dynamically stable disk.

*"Buckingham/Pi Theorem" Considerations*

We can derive the scaling BTFR pre-factor using a Henriksen's interpretation of the Buckingham (or Pi) theorem (Henriksen 2019). Based on the author's nomenclature, there are three physical dimensions: mass $M$, length $L$, and time $T$ (n=3). The associated scaling "galaxy catalog" of measurables is $GC1 \equiv \{G, M_{Dyn}, R_D, V_C\}$ (m=4). Per the Pi theorem, there is only (m-n)=1 dimensionless number required that interrelates these three dimensions, providing the complete set required to completely describe galactic properties. There is one dimensionless number based on the virial theorem that links all parameters with respect to $M_{Dyn}$:

$$\mathcal{V} = 1 = \frac{M_{Dyn} G}{R_D V_C^2}$$

Via a simple recasting and substitution (D=$M_{Dyn}/M_{Bar}$), the scaling version of the baryonic Tully-Fisher relation is produced:

$$V_C^4 = \frac{M_{Dyn}^2 G^2}{R_D^2} = D \frac{ML^6}{L^2 M^2 T^4}(M_{Bar}) = D\left(\frac{L^2}{M}\right)\left(\frac{L^2}{T^4}\right)(M_{Bar}) = \frac{\alpha^2}{\mu_{RD}}(M_{Bar})$$

Taking the analysis further, the galaxy catalog can include two additional parameters; total galactic Energy ($E$) and Angular Momentum ($J$). The expanded catalog is $GC2 \equiv \{G, M_{Dyn}, R_D, V_C, J, E\}$. This requires two additional dimensionless numbers (m-n)=2. For this purpose, the Peebles ($\lambda_P$) and the isothermally constrained Bullock ($\lambda_B$) spin equations suffice with $\lambda_P = \lambda_B$ (Peebles 1971) (Bullock 2001) (Knebe 2011):

$$\lambda_P = \frac{J\sqrt{E}}{M_{Dyn}^{5/2} G} \quad ; \quad \lambda_B = \frac{J}{\sqrt{2} M_{Dyn} V_C R_D}$$

For dimensionless spin coefficient $\lambda$=0.423, the Milky Way's 'dynamic' energy and angular momentum are $E$=1.3x10$^{16}$M$_\odot$km$^2$s$^{-2}$ and $J$=9.8x10$^{31}$M$_\odot$km$^2$s$^{-1}$, respectively.

*Summary*

A proposed scaling model for galactic disks has been presented and compared against the MOND and dark matter frameworks. Power law scaling relations and dimensional analysis are employed to rewrite the functional form of the BTFR in terms of three parameters; disk dynamic surface density/acceleration and mass discrepancy. The role of thermodynamics (galactic virial properties) still has to be understood in shaping these and other galactic scaling relations.

*Acknowledgement*

We thank anonymous referees for assisting in this proposal.